# Metamaterial Thermoelectric Conversion

Takuya Asakura[1, †], Toshinari Odaka[1, †], Ryosuke Nakayama[1], Sohei Saito[1], Shohei Katsumata[1], Takuo Tanaka[*2,3,4,5], and Wakana Kubo[*1]

**Affiliations:**

[1] Division of Advanced Electrical and Electronics Engineering, Tokyo University of Agriculture and Technology, 2-24-16 Naka-cho, Koganei-shi, Tokyo 184-8588, Japan

[2] Metamaterial Laboratory, RIKEN Cluster for Pioneering Research, 2-1 Hirosawa, Wako, Saitama, 351-0198, Japan

[3] Department of Physics, Faculty of Science, Gakushuin University, 1-5-1 Mejiro, Toshima-ku, Tokyo 171-8588, Japan

[4] Innovative Photon Manipulation Research Team, RIKEN Center for Advanced Photonics, 2-1 Hirosawa, Wako, Saitama 351-0198, Japan

[5] Institute of Post-LED Photonics, Tokushima University, 2-1 Mishima, Minami-Jyosanjima, Tokushima 770-8560, Japan

*Correspondence to: w-kubo@cc.tuat.ac.jp, t-tanaka@riken.jp

† These authors contributed equally to this work.

**Abstract:** We propose a thermoelectric device that can produce a thermal gradient even in a uniform-temperature environment. We introduced a metamaterial absorber (MA), which comprised a transparent calcium fluoride layer sandwiched between a silver mirror layer and silver microdisk arrays, at one end of a thermoelectric device made of bismuth antimony telluride. The heating efficiencies of the MA and opposite electrodes of the device were unbalanced; consequently, the Seebeck effect was induced, resulting in electricity generation. We fabricated the MA on a copper electrode loaded on a thermoelectric device, and the device was placed in a uniform-temperature environment at 364 K. The thermal gradient across the device was experimentally measured and was found to be 0.14 K.

**One Sentence Summary:** We propose a metamaterial thermoelectric device that can be installed in uniform-temperature environments where conventional thermoelectric devices cannot be applied.

Thermoelectric devices convert thermal energy into electricity; they can therefore be used to recover waste heat and convert it into a usable energy resource (*1-6*). The Seebeck effect plays a significant role in thermoelectric conversion, in which the amplitude of the temperature gradient across a thermoelectric device is the key to determine the electric voltage generated across it. Since the Seebeck effect converts the temperature gradient across a thermoelectric device to electric voltage, such devices are usually installed only at the locations where large thermal gradients can be expected, such as space probes and wearable electronics, and the installation sites for conventional thermoelectric devices are strongly limited (*7, 8*). Although the Seebeck effect cannot be used under a uniform-temperature environment, there are a lot of such places like as in water, on the road, inside the furnace, etc. Therefore, new techniques that can create temperature gradients even if the devices are surrounded by uniform-temperature environments are in demand. Such technologies can expand the range of installation of thermoelectric devices, thereby contributing to an energy-recycling and energy-saving society.

In this paper, we propose a thermoelectric device loaded with a metamaterial absorber (MA) at the surface of one of its ends to generate electricity even in a uniform-temperature medium. Metamaterials are artificial materials that exhibit extraordinary interaction with electromagnetic waves (*9-11*). Using the large difference in absorptivity between the device edges produced by the metamaterial absorber, the device creates a temperature gradient even the device is basked in uniform and isotropic thermal radiations. Our proposal can break the notion that thermoelectric conversion cannot occur under a uniform-temperature environment. Unlike conventional thermoelectric devices, the metamaterial thermoelectric device can collect and extract thermal energies from the surrounding medium. Thus, these devices will pave the way to recover waste heat existing in the medium, such as air and water.

A metamaterial absorber (MA) consisting of a silver (Ag) film and a Ag disk array sandwiching a calcium fluoride ($CaF_2$) layer was fabricated on a copper electrode (Figure 1(A, B)). Figure 1(C) shows a comparison of the calculated and measured absorption spectra of MA and a blackbody radiation at 364 K. The measured absorption spectrum shows a resonance peak at 6.08 μm attributed to its magnetic resonance mode (*12*), indicating that the MA arrays absorb a part of the thermal radiation emitted from the surrounding medium at 364 K and convert the absorbed energy into local heating (*13-22*).

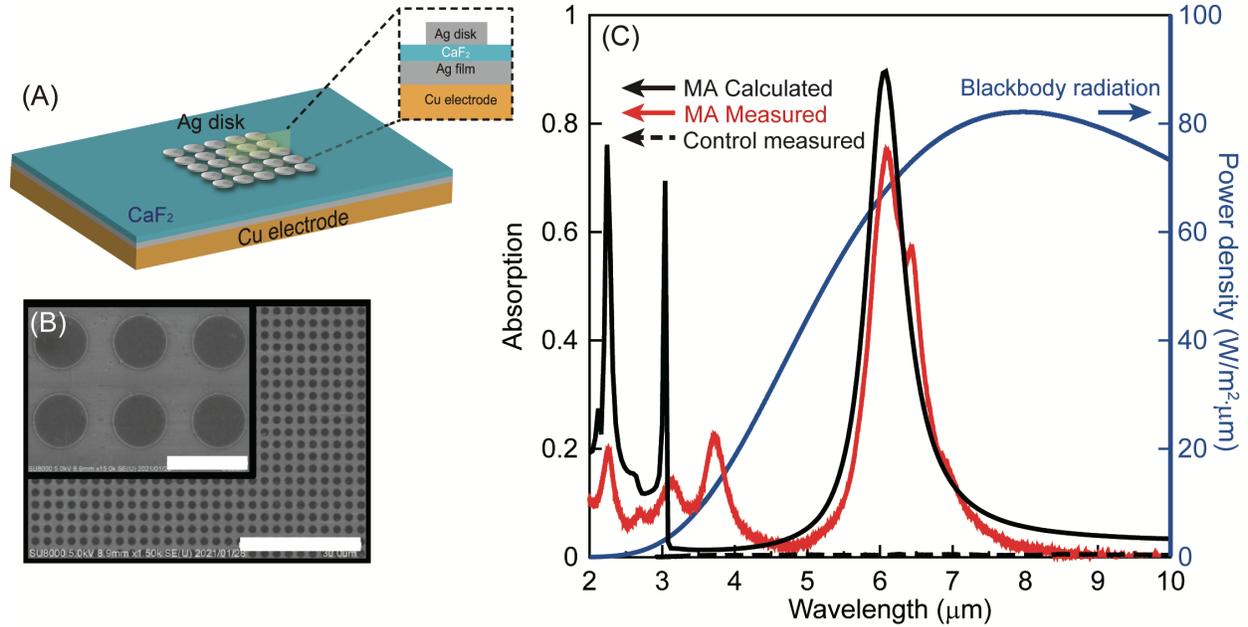

**Fig. 1.** (**A**) Schematic of MA arrays fabricated on a copper electrode. (**B**) A top-view scanning electron microscope (SEM) image of MA arrays. The inset is a magnified image, and the scale bars in the low and high magnified images represent 30 and 3 μm, respectively. (**C**) Comparison of the measured (red) and calculated (black) absorption spectra of the MA, a measured absorption spectrum of the control electrode (dashed black), and a calculated blackbody radiation spectrum (blue) at 364 K.

The MA-fabricated-copper electrode was attached to one end of a p-type bismuth antimony telluride ($Bi_{0.3}Sb_{1.7}Te_3$), provided by TOSHIMA Manufacturing Co., Ltd. A control electrode was attached to the opposite end of the thermoelectric device. The control electrode consisted of $CaF_2$ and Ag layers deposited on a copper electrode. The dashed line in Figure 1(C) is the measured absorption spectrum of the control electrode, indicating that the control electrode does not absorb thermal radiation. The device loaded with the MA fabricated copper electrode is called the MA thermoelectric device (Figure 2(A)). In addition, we prepared a control thermoelectric device loaded with control electrodes at both edges of $Bi_{0.3}Sb_{1.7}Te_3$. To examine the effect of the MA electrode on the thermoelectric performance, we replaced the control electrode of the control device with the MA electrode to model the MA device by retaining the $Bi_{0.3}Sb_{1.7}Te_3$ thermoelectric element of the control device; hence, the difference between the MA device and the control device is only an attachment of the electrode.

The MA thermoelectric device was placed in an electric furnace, and the output voltage generated across the MA thermoelectric device was measured. To eliminate the convection effect on thermoelectric generation, the MA device was capped with a carbon pod (Figure 2(B), and (C)). Figure 2(D) illustrates the dependence of the output voltage on the measured environmental temperatures. The values of output voltage generated across the MA and control device were 18.9 ± 6.7 μV and -1.2 ± 4.4 μV (number of the measured sample: 5), respectively, at the measured environmental temperature of 364 K. The output voltage generated across the MA device was significantly larger than that across the control device. The output voltage generated across the MA device corresponded to an additional temperature gradient of 0.14 K estimated by the measured Seebeck coefficient of the p-type $Bi_{0.3}Sb_{1.7}Te_3$ element (140 µV/K).

In addition, the output voltages generated across these devices were measured at different environmental temperatures (Figure 2(D)). The output voltages generated across the MA thermoelectric device showed significant temperature dependence, whereas those generated on the control device showed no temperature dependence. At every environmental temperature, the thermistors showed a maximum temperature difference of less than ± 0.2 K, hence we concluded that the temperatures at the vicinities of the left and right electrodes were the same. Figures 2(D) and S3 show that the energy absorbed by the MA increases as the environmental temperature increases, indicating an increase in the amount of local heating.

Figure 2(E) shows a comparison between the additional temperature gradients estimated by the measured output voltages and the calculated power densities absorbed by the MA. This indicates that the additional temperature gradients created by the MA electrode originate from thermal radiation absorption by the MA.

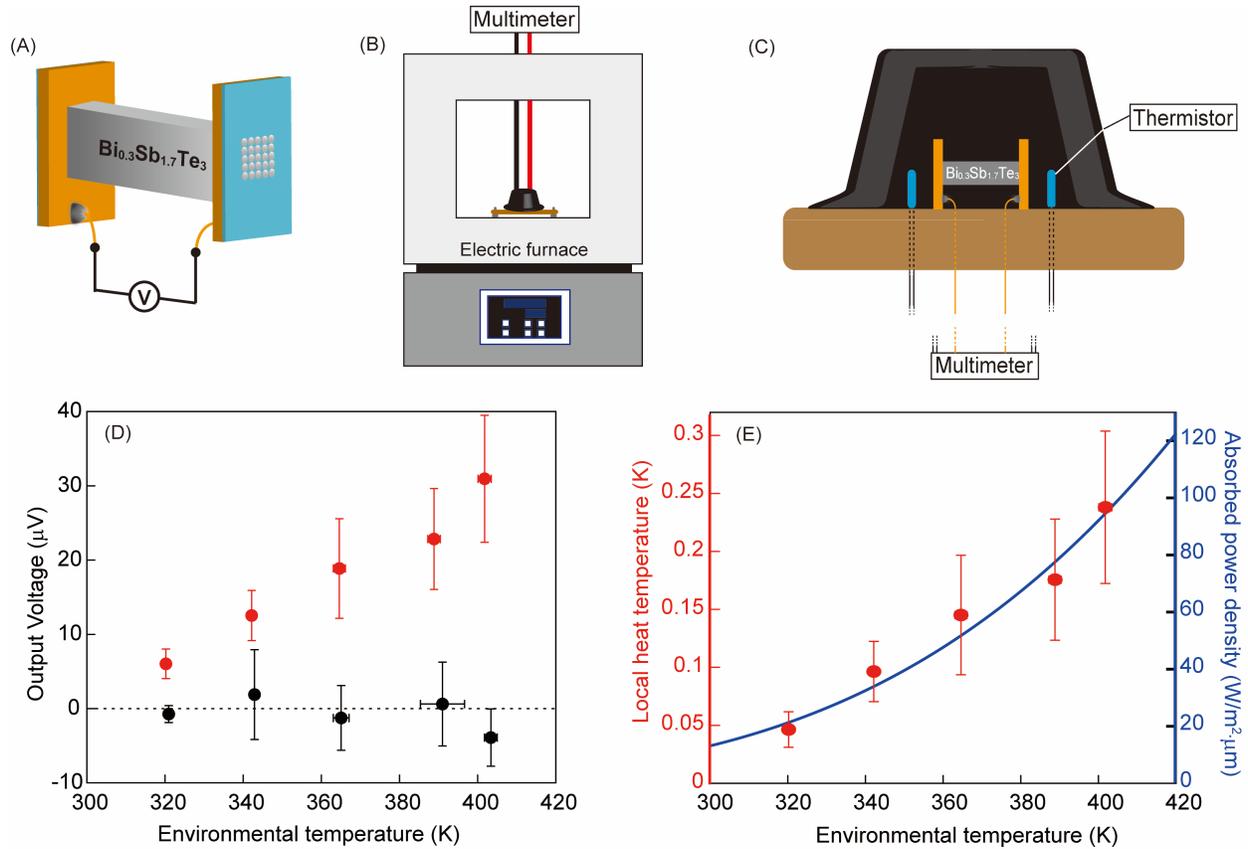

**Fig. 2.** (**A**) Schematic of the $Bi_{0.3}Sb_{1.7}Te_3$ thermoelectric device loaded with the MA electrode. The MA electrode was mounted on one end of the $Bi_{0.3}Sb_{1.7}Te_3$ element, and a control electrode was attached on the other end of the element. (**B**) Experimental setup for thermoelectric measurements. (**C**) Schematic of the MA thermoelectric device and two thermistors capped with a carbon pod. (**D**) Dependence of the output voltages generated on the MA device (red) and a control device (black) on the measured environment temperatures. (**E**) Correlation between the local heat temperatures generated on the MA-$Bi_{0.3}Sb_{1.7}Te_3$ thermoelectric device and the calculated power density absorbed by the MA.

These results indicate the mechanism of metamaterial thermoelectric voltage generation in a uniform-temperature environment. MA absorbs thermal radiation emitted from the surrounding environment and generates local heating due to absorption losses by the MA. The local heat propagates to the $Bi_{0.3}Sb_{1.7}Te_3$ thermoelectric device via a copper electrode, resulting in an additional thermal gradient across the $Bi_{0.3}Sb_{1.7}Te_3$ device and subsequent output voltage generation.

To ascertain our understanding of the driving mechanism of metamaterial thermoelectric conversion, we calculated the local heat temperatures generated on the MA (See "Numerical calculation" for details). The calculated local heat temperature of 0.16 K was obtained on the MA area, which is of a comparable level to the additional temperature gradient estimated by the output voltage generated at the environment temperature of 364 K. In contrast, heat generation corresponding to the temperature of 0.02 K was observed on the surface of the control electrode.

Table S1 presents the calculated and measured local temperatures generated at each element of the MA thermoelectric device. We considered that the local temperatures generated on the $Bi_{0.3}Sb_{1.7}Te_3$ element would not affect the thermoelectric performance of the MA device for the following reasons. The $Bi_{0.3}Sb_{1.7}Te_3$ thermoelectric element has an isotropic shape, meaning that the thermal radiation absorption of $Bi_{0.3}Sb_{1.7}Te_3$ does not induce temperature gradients across the element owing to uniform local heat generation across the element. On the other hand, the local heat generated on the MA electrode creates an anisotropic thermal distribution across the thermoelectric device. This means that the local heat generated on the MA electrode is the key to determining the output voltage across the device, regardless of the amount of temperature generated on the $Bi_{0.3}Sb_{1.7}Te_3$ thermoelectric element. As a proof of this concept, we carried out a thermoelectric simulation.

Figure 3(A) and (B) present simulated temperature difference and electric potential distributions across the MA thermoelectric device and control device. Position X = 0 mm corresponds to the MA electrode surface of the MA device or a left-side electrode surface of the control electrode. The thermal distribution of the MA device in Figure 3(A) showed a temperature difference of 0.14 K between the electrode surfaces, which shows a good agreement with the temperature difference obtained by experiments, 0.14 K. In contrast, the control device indicated that a temperature gradient has not been induced.

The calculated electric potential distributions across the MA thermoelectric device shown in Figure 3(B) showed a potential difference of 19.6 µV, which is comparable to the output voltage experimentally obtained at an environmental temperature of 364 K. Meanwhile, the control electrode did not show a potential difference between the electrodes.

Figure 3(C) presents the time dependence of the potential differences between points a and b (Figure 3(C), upper inset) of the MA thermoelectric device and the control device. A potential difference of 19.6 µV in the MA thermoelectric was observed at 180 min. In contrast, the potential difference observed for the control device is almost negligible, approximately 2.0 nV. Thermoelectric simulation supports our hypothesis that the MA triggers the generation of a thermal gradient across the thermoelectric device placed in uniform-temperature air.

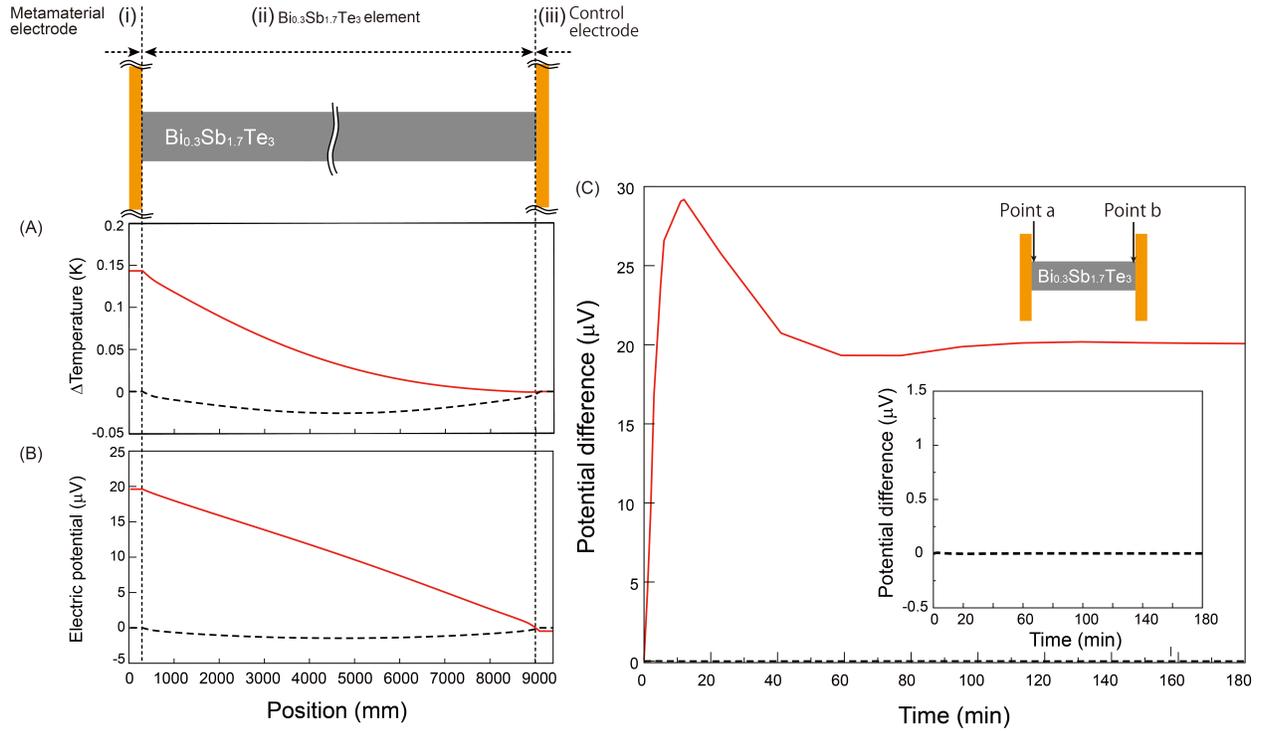

**Fig. 3** (**A**) Temperature difference and (**B**) electric potential distributions of the MA device (red line) and the control device (dashed black line) calculated at 180 min. The area (i), (ii), and (iii) corresponds to the MA electrode, the $Bi_{0.3}Sb_{1.7}Te_3$ element, and the opposite electrode, respectively. (**C**) Calculated time dependence of the potential differences between points a and b (the upper inset) of the MA device (red line) and a control device (dashed black line). Point a is close to the MA electrode for the MA device. The right lower inset is an enlarged potential difference graph of the control device.

In this research, we induced a thermal gradient across a thermoelectric element through the absorptivity control of both electrodes. The unbalanced absorptivity difference between the electrodes produced by metamaterials can lead to a thermal distribution across a certain object or space.

Metamaterial thermoelectric conversion can enhance the thermoelectric performance by increasing the thermal gradient across a thermoelectric device by up to 14% in an environment of 364 K, meaning that metamaterials can enhance the performance of conventional thermoelectric devices installed in a conventional site where a thermal gradient is maintained. It should be noted that the metamaterial thermoelectric conversion cannot drive under uniform-temperature environments without thermal radiation, such as space. This thermal engineering technique can be applied to control thermal distributions in microfluidics, electric circuits, and containers that are filled with uniform-temperature media (*23-26*).

**Acknowledgment:** The authors thank TOSHIMA Manufacturing Co., Ltd. for providing p-type $Bi_{0.3}Sb_{1.7}Te_3$ element.

**Funding:** W. Kubo thanks the Sumitomo Foundation (191029), the JSPS KAKENHI Grant number JP-20K05261, and the JSPS Core-to-Core Program. The authors acknowledge financial support from JST CREST Grant Number JPMJCR1904 and KAKENHI JP-18H03889, Japan.

**Author contributions:** W. K. conceived the idea of this study. T. A., T. O., R. N., S. S., and S. K. performed the fabrication and measurements with support from W.K. and T. T.. T. T. coordinated the numerical calculations. W.K., T. T., and T. A. wrote the manuscript. All authors contributed to the analysis and interpretation of results. W.K. supervised and coordinated the work.

**Competing interests:** Authors declare no competing interests.

**Data and materials availability:** All data and materials except thermoelectric materials provided by TOSHIMA Manufacturing Co., Ltd. are available in the manuscript of the supplementary materials.

**Supplementary Materials:**

Materials and Methods

Figures S1-S5

Table S1

# Supplementary Materials for

## Metamaterial Thermoelectric Conversion

Takuya ASAKURA, Toshinari ODAKA, Ryosuke NAKAYAMA, Sohei SAITO, Shohei KATSUMATA, Takuo TANAKA *, and Wakana KUBO*

Correspondence to: w-kubo@cc.tuat.ac.jp (W. K.), t-tanaka@riken.jp (T. T.)

**This PDF file includes:**

    Materials and Methods
    Supplementary Text
    Figs. S1 to S5
    Table. S1

**Materials and Methods**

Fabrication of metamaterial absorber (MA)

We attached metamaterial arrays to a copper electrode in contact with a thermoelectric device to design a thermal distribution across the device. A 300 μm-thick-copper plate with a size of 4 × 6 mm was thoroughly washed with acetone and pure water with sonication. A 150 nm-thick silver (Ag) layer and a 60 nm-thick calcium fluoride ($CaF_2$) layer were deposited on the copper plate using a thermal evaporator (SVC-7TS, Sanyu Electron Corporation).

To fabricate the MA arrays, hole patterns were drawn on a positive resist layer (ZEP520A, ZEON CORPORATION) formed on the $CaF_2$ and Ag layers deposited on a copper plate using electron beam lithography (JBX-6300FS, JEOL Ltd.), followed by development. The copper electrode area was 4 × 6 $mm^2$. An Ag layer with a thickness of 100 nm was deposited on the patterned resist layer by thermal deposition. Ag disk arrays were obtained after the lift-off process (Figure 1(B)). The diameter and pitch of the Ag disk were 1.8 and 3.0 μm, respectively. These parameters of the MA were employed to achieve an absorption peak at a wavelength of 6 μm, which overlaps with the blackbody radiation spectrum at 364 K (91 °C). The MA arrays were fabricated at the center of the copper electrode with an area of 2.1 × 2.1 $mm^2$ (490,000 units of the MA in the fabricated area).

Preparation of the metamaterial thermoelectric device and a control device

The MA-fabricated-Cu electrode was attached to the edge of a $Bi_{0.3}Sb_{1.7}Te_3$ thermoelectric element by soldering paste. $Bi_{0.3}Sb_{1.7}Te_3$ was utilized in this research since this material exhibits promising thermoelectric performance. $Bi_{0.3}Sb_{1.7}Te_3$ thermoelectric elements with a cross-sectional area and a length of 1 × 2 $mm^2$ and 8.8 mm, respectively, were provided by TOSHIMA Manufacturing Co., Ltd. A control electrode was attached to the opposite end of the thermoelectric element. The difference between the MA and control electrode is the existence of the Ag disk on the $CaF_2$ and Ag layers in case of the MA, meaning that we can control the absorption properties of the MA effectively. For electrical connections, gold wires were connected to each copper electrode using Ag paste.

Optical characteristics measurement of metamaterial absorber

The reflection spectra of the MA arrays, fabricated on a copper electrode, were measured by microscopic Fourier transform infrared spectroscopy (FT/IR-6300, VIRT-3000, JASCO Corporation). The reference spectrum was recorded on a bare copper plate in the vicinity of the MA pattern. For the reflection spectrum of a bare copper surface, a 150 nm-thick Ag layer deposited on bare copper was used as a reference. The absorption spectra (Figure 1(C)) were calculated using the 1- reflectance.

Seebeck coefficient measurement

A control device loaded with control electrodes was used to measure the Seebeck coefficient of the p-type $Bi_{0.3}Sb_{1.7}Te_3$ thermoelectric element. One edge of the device was attached to the surface of a hotplate. The temperature at the rear of the electrodes was monitored by thermistors, and the voltage was measured using a multimeter (DMM-6500, Keithley Instrument).

Thermoelectric measurement

An electric furnace (FUL210FA, ADVANTEC) was used as a heater to create a uniform-temperature environment. The $Bi_{0.3}Sb_{1.7}Te_3$ thermoelectric device capped with a carbon pod with a diameter and a height of 2.5 and 2.5 mm, respectively, was placed on a plastic board at the middle of the furnace chamber. Two thermistors were placed 2 mm away from the electrode surfaces of the thermoelectric device to monitor the environmental temperatures in the vicinity of the electrodes (see Figure 2(C)). The thermistors were calibrated carefully to show the same temperature in the same temperature environment (see "Thermistor calibration", Figure S1).

The entire device was placed on a universal substrate located in the middle of the furnace chamber and capped with a carbon pod. We ensured that the bases of both electrodes were in contact with the universal substrate to avoid any effect on the thermoelectric performance due to the inhomogeneous contact between the electrode and substrate.

To eliminate the convection effect on thermoelectric generation, the MA device was capped with a carbon pod with a diameter and height of 2.5 and 2.5 cm, respectively (Figure 2(B), and (C)). The convection created by the furnace fan winds influenced the thermoelectric performance since it induced a transient thermal gradient across the thermoelectric device. During heating, the electric furnace door was closed, and heating was carried out in a closed chamber.

The MA thermoelectric device was placed in an electric furnace to examine the output voltage generated across the MA thermoelectric device. The output voltage generated on the thermoelectric device was measured using a multimeter. To obtain the output voltage, we averaged the output voltage from 10800 s (180 min) to 11400 s (190 min) after the furnace was started heating (Figure S2). Furthermore, the output voltages were examined with several MA and control electrodes of several different thermoelectrical devices to calculate the average voltages, which indicated that these output voltages were highly reproducible. Five readings of the output voltage and corresponding environmental temperatures inside the furnace were recorded and were represented by error bars (Figure 2(D)).

In addition, we prepared a control thermoelectric device loaded with control electrodes at both ends of $Bi_{0.3}Sb_{1.7}Te_3$. To examine the effect of the MA electrode on the thermoelectric performance, we replaced the control electrode of the control device with the MA electrode to model the MA device by retaining the $Bi_{0.3}Sb_{1.7}Te_3$ thermoelectric element of the control device; hence, the difference between the MA device and the control device is only an attachment of the electrode. This procedure clarified the effect of MA on the device performance because the $Bi_{0.3}Sb_{1.7}Te_3$ thermoelectric element was retained.

Because the $Bi_{0.3}Sb_{1.7}Te_3$ element of the MA device was identical to that of the control device, output voltage measurements of these two devices were carried out, separately.

Thermistor calibration

A thermistor probe (Micro-BetaCHIP thermistor probe, Measurement Specialties, Inc.) was used to measure the environmental temperature. The thermistor is a negative temperature coefficient (NTC) type and has a probe head with a length and a diameter of 3.3 and 0.3 mm, respectively. The thermistor was connected to a temperature sensing circuit that produced an output voltage linear to the environmental temperature.

We utilized the calibrated B values of the thermistor to estimate the environmental temperatures. The B value is an indication of a graph curve that represents the relationship between the resistance and temperature of the thermistor. We measured the resistance of the thermistors at 298 and 353 K to calculate the intrinsic B value for each thermistor. The B values for thermistors 1 (left side) and 2 (right side) were 4288 and 4277, respectively. We placed thermistors 1 and 2 close together at the same position to be exposed to the same temperature in a uniform-temperature environment and calculated the temperature difference between the two thermistors after the environment temperature reached an equilibrium. Figure S1 presents the dependence of the temperature differences between the two thermistors on the environmental setting temperatures. The temperature differences between the two thermistors were in the range of ± 0.2 K at every setting temperature, indicating that the thermistors showed a maximum temperature difference of ± 0.2 K even though the thermistors have the same temperature. Based on these experiments, we concluded that the environmental temperature measured by the left and right thermistors is the same if the temperature difference between the two thermistors is less than ± 0.2 K.

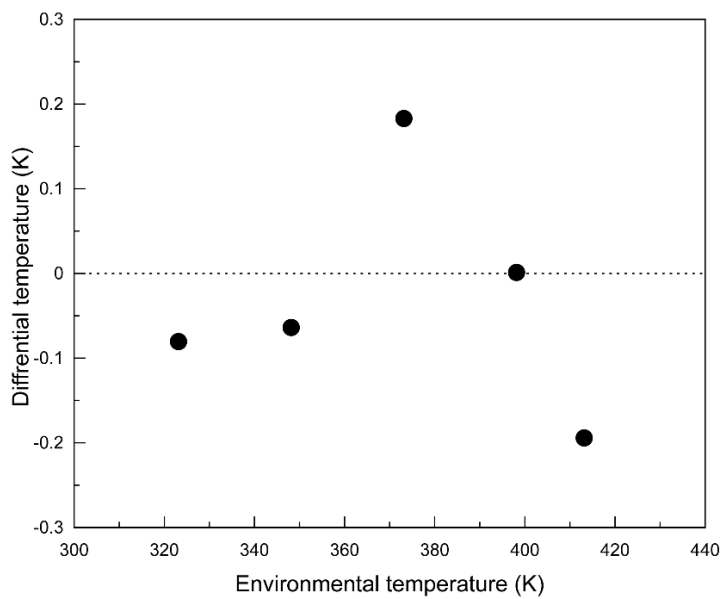

**Fig. S1.** Dependence of the differential temperature between the left and right thermistors on the setting temperatures of an electric furnace. The circle plots present the temperature differences between the left and right thermistors calculated by the calibrated (red) and the catalog (black) B values.

Numerical calculation to estimate local heat temperature

COMSOL Multiphysics software was used to calculate local heat temperature. The calculations were performed at an environmental temperature of 364 K, which was the measured temperature obtained at an electric furnace temperature setting of 373 K. The calculation details are reported in (*1*). The absorbed energies of the MA, corresponding to the heat power density $Q(\mathbf{r})$, can be calculated using the electric field intensity, as described in Eq. (S1).

$$Q(\mathbf{r}) = \frac{\omega}{2} Im(\varepsilon(\omega))\varepsilon_0 |\mathbf{E}(\mathbf{r})|^2, \quad (S1)$$

where $Im$ is the imaginary part, $\varepsilon(\omega)$ is the relative permittivity of the material, $\varepsilon_0$ is the vacuum permittivity, and $|\mathbf{E}(\mathbf{r})|^2$ is the intensity of the electric field. The local heating can be calculated using Eq. (S2) by considering the radiative cooling effect.

$$\rho(\mathbf{r})c(\mathbf{r})\frac{\delta T(\mathbf{r})}{\delta t} = \nabla \kappa(\mathbf{r}) \nabla T(\mathbf{r}) + Q(\mathbf{r}) + Q_{rad}(\mathbf{r}), \quad (S2)$$

where $\rho(\mathbf{r})$ is the mass density, $c(\mathbf{r})$ is the specific heat, and $T(\mathbf{r})$ is the temperature. $\nabla \kappa(\mathbf{r}) \nabla T(\mathbf{r})$ represents the conductive heat flux, and $Q_{rad}(\mathbf{r})$ is the radiative heat flux described by Eq. (S3). We considered the radiative cooling effect of the MA in these calculations, since a good absorber is a good emitter according to Kirchhoff's law (*2-4*). We calculated the local heat temperature using a heat transfer equation that involved the radiative heat flux term $Q_{rad}(\mathbf{r})$ (Eq. (S2)). This equation means that some of the energies absorbed by the MA will be lost as radiative heat flux. The radiative heat flux can be represented using the Stefan–Boltzmann equation (Eq. (S3)),

$$Q_{rad}(\mathbf{r}) = \varepsilon \sigma A (T(\mathbf{r}) - T_{surr})^4, \quad (S3)$$

where $\varepsilon$ is the emissivity, $\sigma$ is the Stefan–Boltzmann constant, $A$ is the area of the absorber, $T(\mathbf{r})$ is the surface temperature of the MA, and $T_{surr}$ is the surrounding environmental temperature.

We concluded that the radiative heat fluxes certainly exist in our system; however, its effect is negligible because of two reasons. First, radiative cooling will occur at an atmospheric window of 8–13 μm, which is outside of the MA resonance utilized in this study. Second, our experiments were conducted in a closed carbon pod placed in a furnace. The walls of the carbon pod and furnace obstruct the radiation cooling. In our case, the radiative heat flux emitted from the MA surface is given to the air in the vicinity of the MA electrode. The temperatures of the MA electrode and surrounding air are approximately the same, resulting in a slight radiation heat flux according to Eq. (S3). In contrast, the absorbed energy of the MA is converted into local heating, leading to effective conductive heat propagation to the copper electrode owing to its high thermal conductivity, which results in an additional thermal gradient across the $Bi_{0.3}Sb_{1.7}Te_3$ thermoelectric device.

To estimate the local heat temperature, we obtained the dependence of the local temperatures on the number of units of the MA and control models (Figure S4), and the logarithmic fit to the calculated data as per Eqs. (S4) and (S5). We previously obtained these equations in (*1*).

$$T_{MA} = 0.0056 \ln(x) + 0.0822 \qquad \text{Eq. (S4)}$$

$$T_{control} = 0.0007 \ln(x) + 0.0111 \quad \text{Eq. (S5)}$$

As there are 490,000 units of the MA in the fabricated area, the calculated local heat temperature of 0.16 K was obtained on the MA area, which is of a comparable level to the additional temperature gradient estimated by the output voltage generated at the environment temperature of 364 K.

Thermoelectric simulation

In the thermoelectric simulation, the calculated power density, which corresponds to the electromagnetic energy density absorbed by the material, is given to the MA and control electrode every second to simulate the thermal distribution across the MA thermoelectric device by considering local heat generation at each material. The power density is the total energy absorbed by the material over given area and can be expressed in units of W/m³. These settings are required because all materials consisting of the device generate local heat depending on its absorption properties. This means that the thermal distribution across the thermoelectric device was formed depending on the local heat temperatures generated on each element. The initial temperature of the thermoelectric simulation was 364.15 K, which is equal to the measured environmental temperature.

In the simulation model, the metamaterial thermoelectric device was surrounded by an airbox with a size of 2.5 × 2.5 × 2.5 cm³, which corresponded to the air space inside the carbon pod. Furthermore, to allow heat transfer from the inside of the airbox to the outside, a larger airbox with a size of 50 × 50 × 50 cm³ was placed to cover the inner airbox.

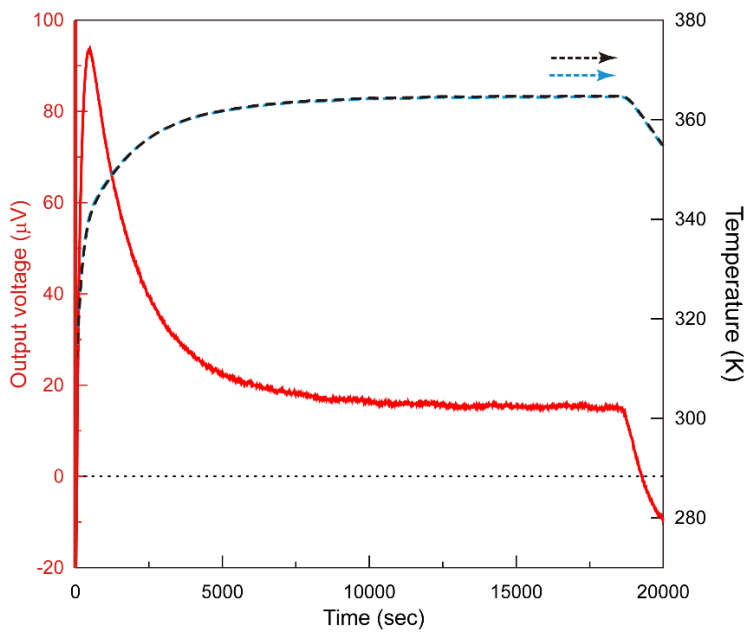

**Fig. S2.** Time dependence of the output voltage generated across the MA thermoelectric device. The red line presents time dependence of the output voltage generated on the MA device, and the dashed lines present the environmental temperatures at the left electrode (dashed black) and at the right electrode (dashed blue) measured by the left and right thermistors. Electric furnace heating with a set temperature of 373 K was started at 0 s and was maintained until the set temperature was achieved at 18500 s. At 18500 s, electric furnace heating was stopped.

The temperature graphs indicate that the measured temperature inside the furnace reached an equilibrium approximately 60 min after the furnace was started and that the temperature difference between the thermistors placed in front of both electrodes was less than 0.2 K. The temperature difference between the left and right thermistors was less than ± 0.2 K when the two thermistors were exposed to the same temperature (see SI Calibration of thermistors). Therefore, we concluded that the environmental temperature at the left and right thermistors is uniform if the temperature difference between the two thermistors is less than 0.2 K. The output voltage was almost constant 120 min after the furnace was started, and we confirmed that the output voltage was stable until the furnace was stopped.

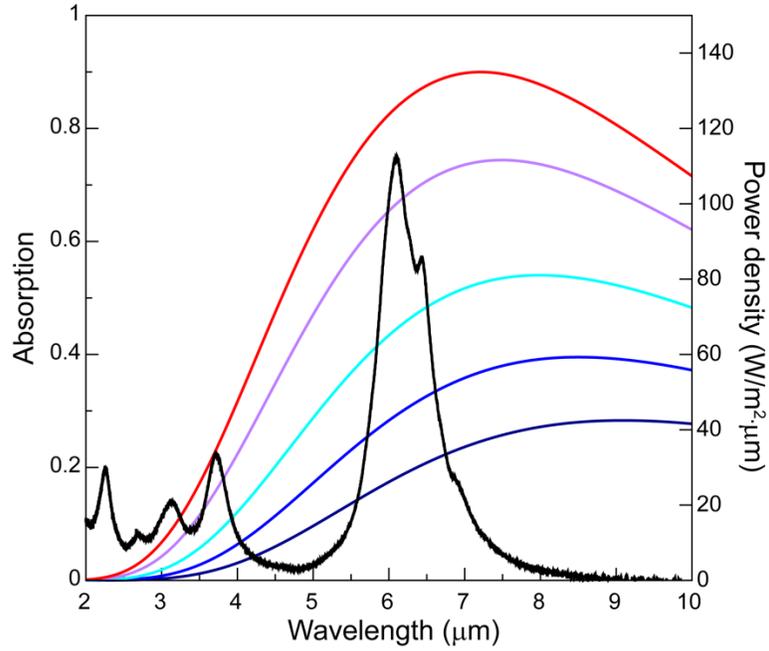

**Fig. S3.** Comparison of the measured absorption spectrum of the MA (black line) and the calculated blackbody radiation spectra at 320 (48 °C, dark blue), 342 (70 °C, blue), 364 (91 °C, light blue), 388 (118 °C, purple), and 402 K (130 °C, red).

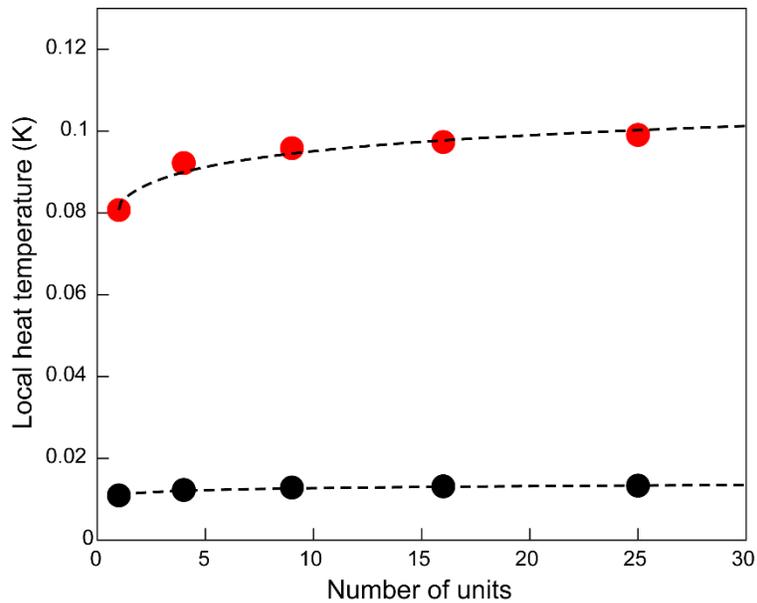

**Fig. S4.** Dependence of the local temperatures on the number of the units in an array. Calculated local heating temperatures as a function of the number of the MA (red) and control (black) units. The dashed lines indicate logarithmic fits to the data.

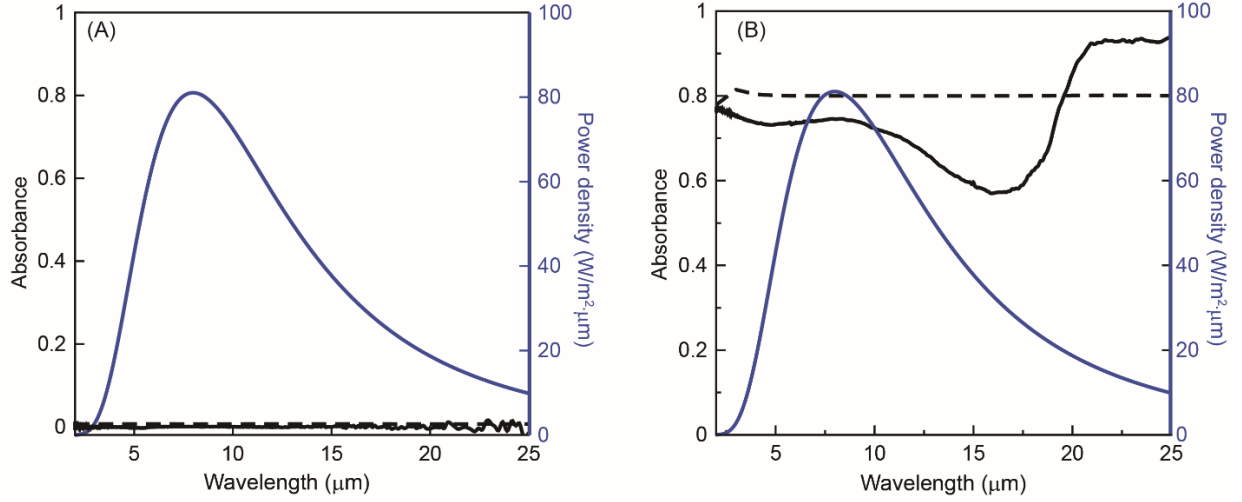

**Fig. S5.** Measured (solid lines) and calculated (dashed lines) absorption spectra of (**A**) a bare copper and (**B**) $Bi_{0.3}Sb_{1.7}Te_3$ element. The blue lines in (A, B) show the blackbody radiation spectra at 364 K.

The copper element showed a slight absorption in the range of 2–25 μm, while the $Bi_{0.3}Sb_{1.7}Te_3$ element exhibited a broad absorption. There is a large difference between the measured and calculated absorption spectra of the $Bi_{0.3}Sb_{1.7}Te_3$ element because we considered bismuth telluride ($Bi_2Te_3$) material properties to estimate the absorption spectrum of the $Bi_{0.3}Sb_{1.7}Te_3$. Furthermore, we believe that the surface roughness of the $Bi_{0.3}Sb_{1.7}Te_3$ element might enhance its apparent absorption. However, we considered that these discrepancies between the measured and calculated absorption spectra are not critical for considering the metamaterial thermoelectric mechanism.

The $Bi_{0.3}Sb_{1.7}Te_3$ thermoelectric element has an isotropic shape, meaning that the thermal radiation absorption of $Bi_{0.3}Sb_{1.7}Te_3$ does not induce temperature gradients across the element owing to uniform local heat generation across the element. On the other hand, the local heat generated on the MA electrode creates an anisotropic thermal distribution across the thermoelectric device. This means that the local heat generated on the MA electrode is the key to determining the output voltage across the device, regardless of the amount of temperature generated on the $Bi_{0.3}Sb_{1.7}Te_3$ thermoelectric element. As a proof of this concept, we carried out a thermoelectric simulation.

We determined the dependence of the local temperature on the number of units (Eqs. (S6, 7)).

$$T_{copper} = 4 \times 10^{-5} \ln(x) + 0.00005 \text{ Eq. (S6)},$$

$$T_{Bi_{0.3}Sb_{1.7}Te_3} = 0.0234 \ln(x) + 0.018 \text{ Eq. (S7)}.$$

**Table S1**. Calculated and measured local temperatures at each element consisting of the MA thermoelectric device.

| Element | Calculated local temperature (K) | Measured temperature (K) |
|---|---|---|
| MA | 0.16 | 0.14 [*1] |
| MA electrode (except MA area, with 60 nm-thick-$CaF_2$ and 150 nm-thick-Ag layers), and Control electrode (with 60 nm-thick-$CaF_2$ and 150 nm-thick-Ag layers) | 0.02 | - |
| Bare copper (Rear side of an electrode) | negligible | - |
| $Bi_{0.3}Sb_{1.7}Te_3$ | 0.34 | 0.13 [*2] |

*1 Estimated by the measured Seebeck coefficient.
*2 Measured temperature using thermistors.